\documentclass[aps,prb,twocolumn,letterpaper,superscriptaddress,showpacs]{revtex4-1}
\usepackage{keyval}%
\usepackage{graphicx}
\usepackage{dcolumn}
\usepackage{bm}
\usepackage{color}
\usepackage{amsmath, amssymb}
\usepackage{soul}

\begin{document}

\title{Hidden mechanism for embedding the flat bands of Lieb, kagome, and checkerboard lattices in other structures}
\author{Chi-Cheng Lee}
\affiliation{Institute for Solid State Physics, The University of Tokyo, 5-1-5 Kashiwanoha, Kashiwa, Chiba 277-8581, Japan}%
\author{Antoine Fleurence}
\affiliation{School of Materials Science, Japan Advanced Institute of Science and Technology (JAIST), 1-1 Asahidai, Nomi, Ishikawa 923-1292, Japan}
\author{Yukiko Yamada-Takamura}
\affiliation{School of Materials Science, Japan Advanced Institute of Science and Technology (JAIST), 1-1 Asahidai, Nomi, Ishikawa 923-1292, Japan}
\author{Taisuke Ozaki}
\affiliation{Institute for Solid State Physics, The University of Tokyo, 5-1-5 Kashiwanoha, Kashiwa, Chiba 277-8581, Japan}%
\date{\today}

\begin{abstract}
The interplay of hopping parameters that can give rise to flat bands in consequence of quantum interference in electronic, photonic, and other interesting materials
has become an extensively studied topic. Most of the recognized structures having flat bands are the lattices that can be 
understood by the mathematical theory of line graphs, such as the Lieb, kagome, and checkerboard lattices. Here, we demonstrate that the 
structures that can realize the same kind of flat bands given by those well-known lattices hosting exotic quantum phases are more flexible. 
The flat bands belonging to the recognized structures can be ideally embedded into the new structures that cannot be considered as the original ones 
in terms of a unitary transformation. 
The uncovered mechanism enriches the understanding of physics behind the localized quantum states and broadens the choice of materials that can be used 
for designing electronic and photonic devices from the zero band dispersion.
\end{abstract}

\maketitle

\section{Introduction}

Much attention has been paid to the studies of massless and infinitely heavy quasiparticles in condensed matter\cite{RevModPhys.81.109,
RevModPhys.90.015001,Parameswaran2013,BERGHOLTZ2013,Liu2014,Derzhko2015,Leykam2018}. As indicated by the names, 
the massless quasiparticles, such as massless Dirac and Weyl fermions\cite{RevModPhys.81.109,RevModPhys.90.015001}, 
can move relativistically in contrast to the infinitely heavy fermions with divergently large effective masses\cite{Parameswaran2013,BERGHOLTZ2013,Liu2014,Derzhko2015,Leykam2018}. 
These two seemingly exclusive behavior could have interesting connections. Perhaps the most well-known example is the Chern number
connecting the dispersionless Landau level with the Weyl point, which is the magnetic monopole of Berry flux, for charactering their topology\cite{RevModPhys.90.015001}.
Both types of quasiparticles can also exist simultaneously in a band structure, for example, 
the one given by the Lieb or kagome lattice\cite{PhysRevLett.62.1201,PhysRevB.62.R6065,PhysRevB.99.125131}. 
Another focus of the flat-band physics is on the destructive quantum interference of localized energy eigenstates existing owing to the energy degeneracy in the entire Brillouin zone, 
which is accessible by a simple tight-binding model, and therefore is of general interest\cite{Parameswaran2013,BERGHOLTZ2013,Liu2014,Derzhko2015,Leykam2018}. 
Early studies of such simple lattice models can be traced back to the realization of localization of electronic wave functions in a dice lattice\cite{PhysRevB.34.5208}, 
finding an ideal model for understanding ferromagnetism\cite{mielke1991ferromagnetic,mielke1991ferromagnetism,mielke1992exact,PhysRevLett.69.1608,Tasaki1998},
and making connection of the lattice structures with line graphs\cite{mielke1991ferromagnetic,mielke1991ferromagnetism,mielke1992exact,PhysRevLett.69.1608,Tasaki1998}. 
The flat-band physics has become an extensively studied topic in a variety of materials, such as the electronic, photonic, 
and cold-atom systems\cite{PhysRevLett.106.236802,PhysRevLett.106.236803,PhysRevLett.106.236804,PhysRevLett.107.146803,PhysRevLett.99.070401,PhysRevB.77.235107,
PhysRevB.82.184502,PhysRevLett.84.143,peotta2015superfluidity,PhysRevB.81.113104,PhysRevB.85.205128,PhysRevLett.114.245503,
PhysRevLett.114.245504,PhysRevB.93.075126,Xia2016,Zong2016,PhysRevA.80.063603,PhysRevB.84.115136,PhysRevA.83.063601,PhysRevB.99.045107}.

To experimentally realize flat electronic bands in real materials, on the other hand, is still challenging and requires more theoretical 
investigations\cite{C8TA02555J,C8NR08479C,su2018prediction,jiang2019lieb}.
Very recently, the flat bands have been directly observed
by angle-resolved photoemission spectroscopy experiments in Fe$_3$Sn$_2$ kagome lattices\cite{PhysRevLett.121.096401} and bilayer graphene\cite{Marchenkoeaau0059}.
Existence of flat bands given by the kagome lattice was also discussed in the recent paper reporting the results of scanning tunneling microscopy 
and scanning tunneling spectroscopy  measurements on twisted multilayer silicene formed on Ag(111)\cite{Lieaau4511}.
The progress has promised to realize many exotic quantum phenomena, for example,
the high-temperature superconductivity associated with the infinitely large density of states of the flat bands\cite{PhysRevB.83.220503,Bistritzer12233,Marchenkoeaau0059},
in real materials soon. 
There is another boost that the ideal kagome bands, which host exotic quantum phases, can also be exhibited in a lattice distinct from the original kagome lattice, 
and the equivalency is verifiable by a unitary transformation\cite{PhysRevB.99.100404}. Along the line of broadening the choice of real flat-band materials with tunability
while still keeping certain exotic quantum phenomena proposed in a simple model, it is interesting to ask whether there still exists physics that has not been revealed so far,
especially so given that many elegant combinations of tight-binding parameters can give rise to the flat bands\cite{JPSJ.72.2015,JPSJ.74.393,JPSJ.74.1918}.

Since the Hamiltonian and its corresponding eigenvalue equation of a system hosting flat bands contain all of the needed ingredients
for revealing the $k$-independent energy eigenvalues, we will demonstrate that the same eigenvalue equation giving rise to 
a flat-band solution, and therefore the same kind of quantum interference, can be embedded into the eigenvalue equation of a new system. 
This new system cannot be obtained by a unitary transformation from the original Hamiltonian because the band structures are in general
different from each other, neither can be directly revealed by the line-graph theory since the embedding is hidden and is only possible under certain conditions. 
We will show that the nature of the flat band existing in the original system is essential for such ideal embedding 
and will introduce a condition that is easy to satisfy by tuning the property of an adatom, such as its height and species, 
in three different lattices deformed from the commonly studied Lieb, kagome, and checkerboard lattices, which will be discussed in detail
in Sec.~\ref{sec:embedding}.

\section{Lattices hosting flat bands and their embedding}
\label{sec:embedding}

In this Section, we will introduce three lattices that can host flat bands by just considering first-neighbor hopping and then
demonstrate how they can be embedded into new structures via deformation and the additon of adatoms. The adatom could be different 
from the original atomic species or the same as the original one. A condition that is important for such embedding 
will be discussed for all the examples. 

\subsection{Lieb lattice}
\label{sec:lieb}

\begin{figure}[tbp]
\includegraphics[width=1.00\columnwidth,clip=true,angle=0]{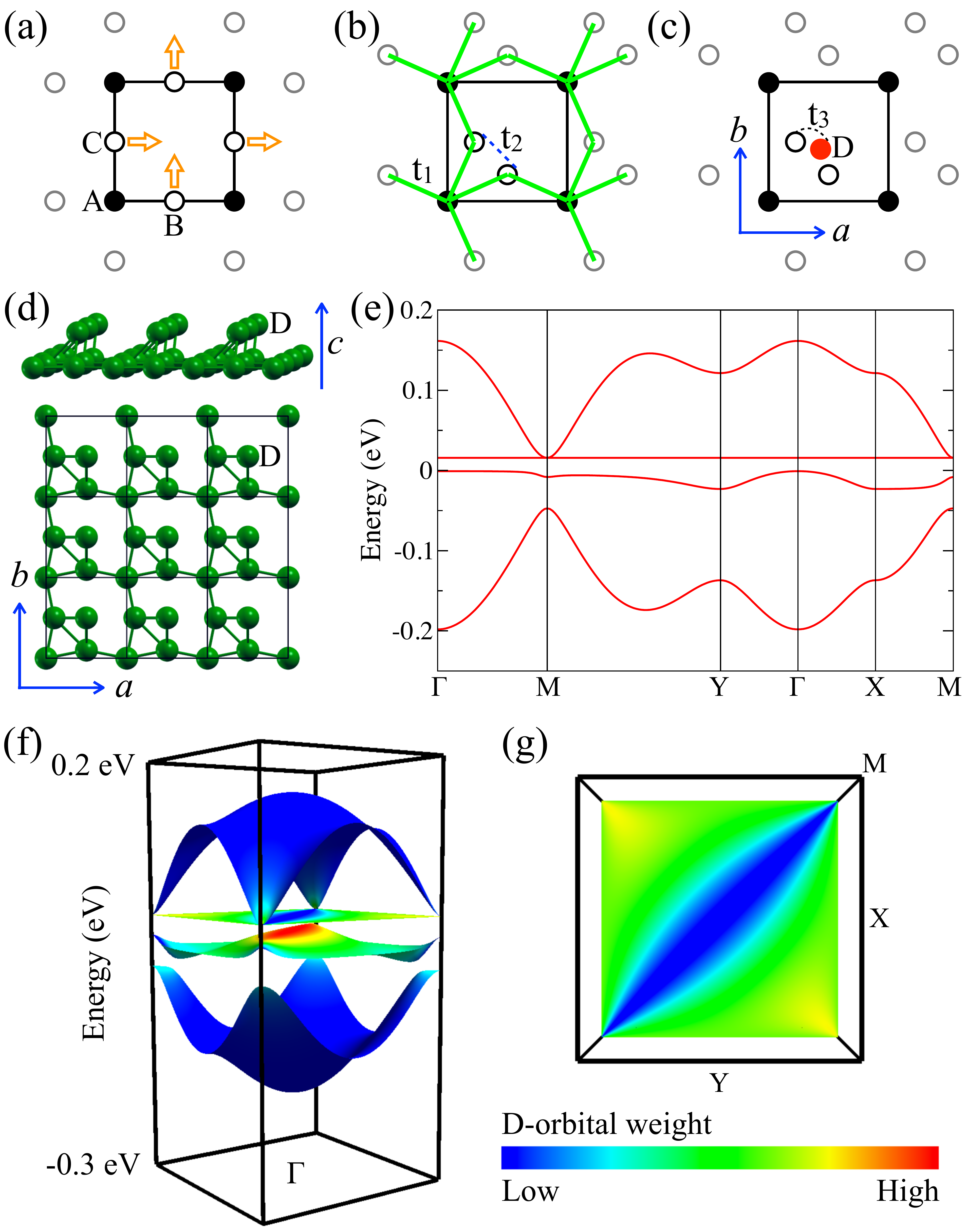}
\caption{(a) Lieb lattice. (b) Distorted Lieb lattice formed by the displacement indicated by the arrows shown in (a). 
(c) An adatom (red circle) is added in (b). (d) A structure composed of H atoms locating at A:(0, 0), B:(0.5, 0.1), C:(0.1, 0.5), and D:(0.5, 0.5)
with the lattice constant $a = 7.5$ \AA. (e) Band structure of (d) for the case of D-atom height at 3.087 \AA~with
$\Gamma$:(0, 0), M:(0.5, 0.5), X:(0.5, 0), and Y:(0, -0.5). 
(f) Three-dimensional band structure with D-orbital contribution. (g) Top view of the flat band with D-orbital weight. 
}
\label{fig:fig1}
\end{figure}

The Lieb lattice having one orbital, say $s$ orbital, per atom with the nearest-neighbor hopping integral, $t_1$, is shown in Fig.~\ref{fig:fig1} (a). 
The site energies of the orbitals locating at A, B, and C sites are denoted as $\epsilon_A$, $\epsilon_B$, and $\epsilon_B$, respectively.
Having these tight-binding parameters, the band structure can be obtained by solving the eigenvalue problem:
\begin{align}
& \epsilon_A C_A + t_1(1+K_a^*)C_B + t_1(1+K_b^*) C_C = \lambda C_A \label{eq:eq1} \\
& t_1(1+K_a) C_A + \epsilon_B C_B + 0 = \lambda C_B \label{eq:eq2} \\
& t_1(1+K_b) C_A + 0 + \epsilon_B C_C = \lambda C_C, \label{eq:eq3}
\end{align}
where $C_A$, $C_B$, $C_C$, $K_a$, and $K_b$ denote the eigenvector coefficients, 
$e^{i2\pi k_a}$, and $e^{i2\pi k_b}$, respectively. The eigenvalue $\lambda$ is in general $k$-dependent and should vary with
the given $k$ point, ($k_a$, $k_b$). However, there exists a $k$-independent solution, $\lambda = \epsilon_B$, at arbitrary $k$ points, 
which is what we call a flat band. 

We now introduce a new structure that can be obtained by displacing the B and C atoms along the $b$ and $a$ directions, respectively. 
Such an arrangement does change the system, for example, the hopping integral between orbital B and orbital C becomes
non-negligible, as denoted by $t_2$ in Fig.~\ref{fig:fig1} (b). The idea to design a material that still behaves as
the Lieb lattice is to add one additional atom, whose species, height, and in-plane position are in principle tunable, into the system.
To realize the same kind of flat band revealed in the Lieb lattice, we shall restrict orbital D can only hop to the B and C orbitals 
via the hopping parameter $t_3$, as shown in Fig.~\ref{fig:fig1} (c). The new equations for the eigenvalue problem become
\begin{align}
& \epsilon_A C_A + t_1(1+K_a^*)C_B + t_1(1+K_b^*) C_C + 0 = \lambda C_A \label{eq:eq4} \\
& t_1(1+K_a) C_A + \epsilon_B C_B + t_2 C_C + t_3 C_D  = \lambda C_B \label{eq:eq5} \\
& t_1(1+K_b) C_A + t_2 C_B + \epsilon_B C_C + t_3 C_D = \lambda C_C \label{eq:eq6} \\
& 0 + t_3 C_B + t_3 C_C + \epsilon_D C_D = \lambda C_D. \label{eq:eq7}
\end{align}
We will show step by step that the flat band of the Lieb lattice can be embedded into this newly introduced structure under one condition based on
the flat-band nature revealed in the original Lieb lattice.

\begin{table}[tbp]
\caption{Tight-binding parameters for the deformed Lieb (L) and kagome (K) structures 
shown in Fig.~\ref{fig:fig1} (d) and Fig.~\ref{fig:fig2} (d), respectively, in unit of meV.
The parameters for the deformed checkerboard (C) structure shown in Fig.~\ref{fig:fig3} (c) is in unit of $t$.
}
\label{table:table1}%
\begin{tabular}{ccccccccc}
\hline\hline
                & $\epsilon_A$ & $\epsilon_B$ & $\epsilon_D$ & $\epsilon_E$ & $t_1$ & $t_2$ & $t_3$ & $\lambda_{flat}$ \\ \hline
L  & -7.6 & -7.2 & -0.7 & & -62.7 & -23.1 & -19.6 & 15.9 \\
K  & -29.3 & -29.3 & -27.7 & & -38.6 & -38.4 & -66.3 & 86.3 \\ 
C & 0    &  0 &  0 & -0.0125 & -1 & -0.8  & -1.5 & 2.8 \\  \hline\hline
\end{tabular}%
\end{table}

First, Eq.~\ref{eq:eq7} can be rescaled by a factor of $t_2/t_3$: 
\begin{align}
t_2 C_B + t_2 C_C + t_2(\epsilon_D-\lambda)/t_3 C_D = 0. \label{eq:eq8}
\end{align}
One can find that Eqs.~\ref{eq:eq5} and \ref{eq:eq6} can be restored to Eqs.~\ref{eq:eq2} and \ref{eq:eq3}, 
respectively, under the condition:
\begin{align}
t_2(\epsilon_D-\lambda)/t_3 = t_3, \label{eq:eq9}
\end{align}
that is, Eq.\ref{eq:eq5} - Eq.\ref{eq:eq8} and Eq.\ref{eq:eq6} - Eq.\ref{eq:eq8}:
\begin{align}
& t_1(1+K_a) C_A + (\epsilon_B-t_2) C_B + 0  = \lambda C_B \label{eq:eq10}\\
& t_1(1+K_b) C_A + 0 + (\epsilon_B-t_2) C_C  = \lambda C_C. \label{eq:eq11}
\end{align}
By comparing Eqs.~\ref{eq:eq4}, \ref{eq:eq10}, and \ref{eq:eq11} with Eqs.~\ref{eq:eq1}, \ref{eq:eq2}, and \ref{eq:eq3},
the effect of the deformation is to shift the site energy of B and C orbitals from $\epsilon_B$ to $\epsilon_B-t_2$. 
Therefore, the flat band is still preserved with a new value of energy, $\lambda=\epsilon_B-t_2$. It should be noted that 
Eq.~\ref{eq:eq9} cannot be satisfied in general for arbitrary $k$ points since the eigenvalue $\lambda(\vec{k})$ is 
$k$-dependent while the site energy and hopping parameters are not. Therefore, 
this embedding is only possible because of the nature of the flat band, whose energy is a $k$-independent constant,
already revealed in the original Lieb lattice.

Our derivation shows that a perfect flat band can appear once Eq.~\ref{eq:eq9} is satisfied by
the parameters given in a tight-binding Hamiltonian, where we assume the physics is dominated by short-range hopping. 
An example for the parameters can be obtained by performing first-principles calculations with the adoption
of atomic orbitals as the basis. The H atom with one $s$ orbital is then adopted and 
deployed into the structure shown in Fig.~\ref{fig:fig1} (d). The atomic radius is chosen as 5 Bohr, and
the lattice constant is set to 7.5 \AA~to avoid long-range hopping. Expectedly, the band structure highly depends on the D-atom height
measured from the plane composed of A, B, and C atoms. When the height reaches 3.087 \AA, Eq.~\ref{eq:eq9} is satisfied and a flat band is revealed as shown 
in Figs.~\ref{fig:fig1} (e) and (f). The three-dimensional band structure is generated using FermiSurfer\cite{KAWAMURA2019197}.
The corresponding tight-binding parameters can be obtained via a unitary transformation from the Bloch states to 
the Wannier-function basis and are listed in Table~\ref{table:table1}, where the condition, Eq.~\ref{eq:eq9}, is perfectly satisfied.
In addition, the circumstance shown in Fig.~\ref{fig:fig1} (b) can also be realized by a peculiar orbital order without
the real-space distortion, as long as the hopping behavior is effectively the same.

\subsection{Kagome lattice}
\label{sec:kagome}

The second example is the structure that has been discussed elsewhere, namely the coloring-triangle lattice\cite{PhysRevB.99.100404}.
In the following discussion, we will call the lattice bitriangular lattice since we will focus on the flat band that can be 
realized by considering all relevant nearest-neighbor hopping parameters and site energies. 
What we will demonstrate is that the flat band of the kagome lattice can be embedded in the bitriangular lattice under the same condition 
that has been illustrated for the Lieb lattice. 

\begin{figure}[tbp]
\includegraphics[width=1.00\columnwidth,clip=true,angle=0]{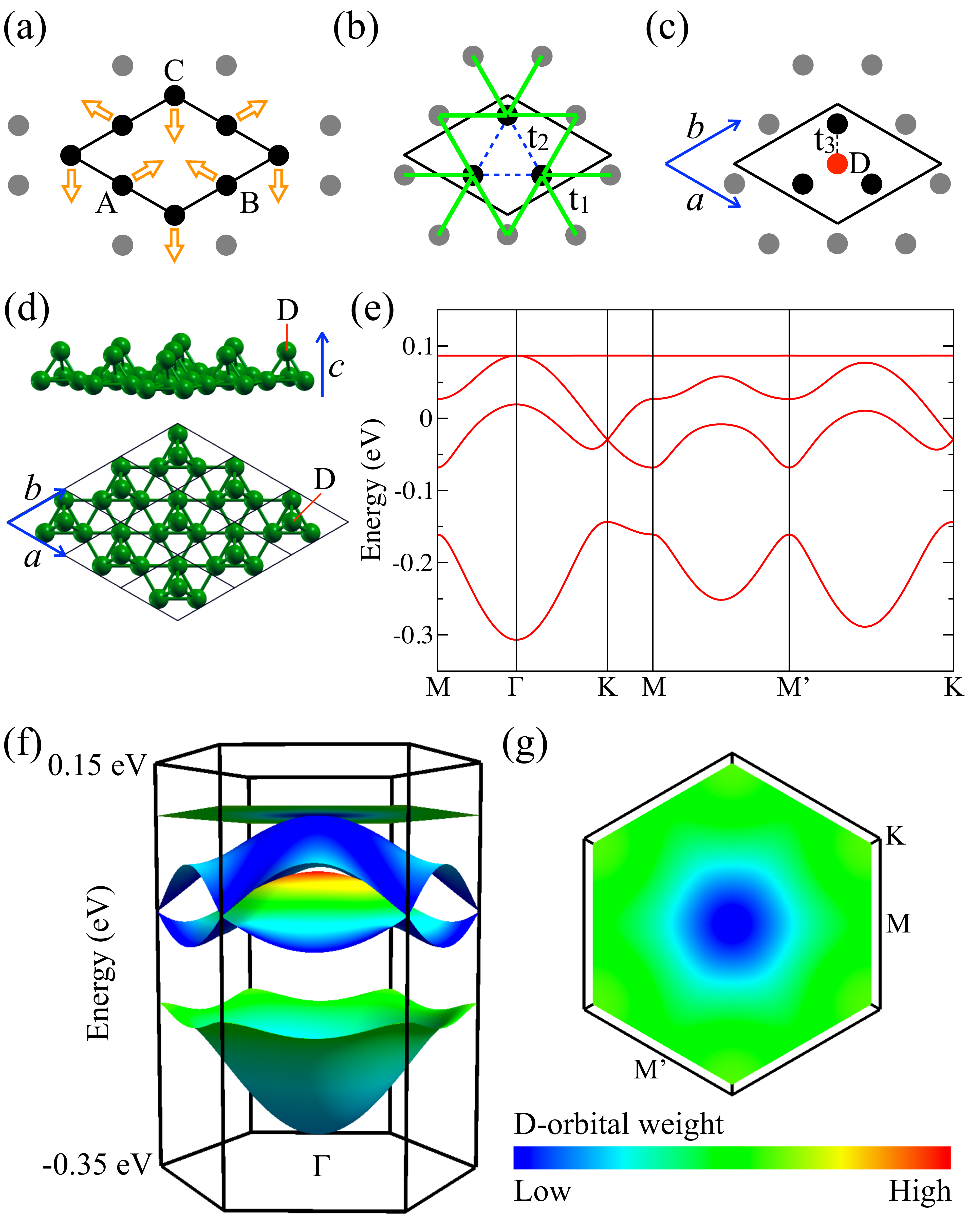}
\caption{ (a) Kagome lattice. (b) Triangular lattice formed by the displacement indicated by the arrows shown in (a). 
(c) Bitriangular lattice formed by adding an adatom (red circle). 
(d) A structure composed of H atoms locating at A:(1/2, 1/6), B:(5/6, 1/2), C:(1/6, 5/6), and D:(1/2, 1/2)
with the lattice constant $a = 7$ \AA. (e) Band structure of (d) for the case of D-atom height at 2.995 \AA~with
the symbols defined as $\Gamma$:(0, 0), K:(1/3, 2/3), M:(1/2, 1/2), and M':(0, -1/2).
(f) Three-dimensional band structure with D-orbital contribution. (g) Top view of the flat band with D-orbital weight. 
}
\label{fig:fig2}
\end{figure}

The band structure of the kagome lattice with the site energy $\epsilon_A$ and 
the nearest-neighbor hopping integral $t_1$ can be obtained by diagonalizing the Hamiltonian:
\begin{align}
\begin{bmatrix}
 \epsilon_A & t_1(K_a^*+K_b^*) & t_1K_b^*(1+K_a)  \\
 t_1(K_a+K_b) & \epsilon_A & t_1K_a(1+K_b^*)  \\
 t_1K_b(1+K_a^*) & t_1K_a^*(1+K_b) & \epsilon_A  
\nonumber
\end{bmatrix}
, 
\end{align}
where the flat band is at the energy $\epsilon_A-2t_1$. After displacing the atoms following the arrows shown in Fig.~\ref{fig:fig2} (a),
a triangular lattice is formed. In this new structure, the hopping parameter $t_2$ indicated in 
Fig.~\ref{fig:fig2} (b) is no longer negligible and should be equal to $t_1$. To allow the same mechanism that gives rise to the flat band in the kagome lattice 
functioning in the triangular lattice, an adatom D, which forms another triangular lattice and can hop to A, B, and C atoms via $t_3$, 
is introduced as shown in Fig.~\ref{fig:fig2} (c). The new Hamiltonian becomes
{\footnotesize
\begin{align}
\begin{bmatrix}
 \epsilon_A & t_2+t_1(K_a^*+K_b^*) & t_2+t_1K_b^*(1+K_a) & t_3 \\
 t_2+t_1(K_a+K_b) & \epsilon_A & t_2+t_1K_a(1+K_b^*) & t_3 \\
 t_2+t_1K_b(1+K_a^*) & t_2+t_1K_a^*(1+K_b) & \epsilon_A & t_3 \\
 t_3 & t_3 & t_3 & \epsilon_D 
\nonumber
\end{bmatrix}
. 
\end{align}
}

One can apply the same mathematics as done for the Lieb lattice by rescaling the last row of the above matrix with the factor 
of $t_2/t_3$ and imposing the condition given by Eq.~\ref{eq:eq9}. The effective Hamiltonian for the flat band becomes
\begin{align}
\begin{bmatrix}
 (\epsilon_A-t_2) & t_1(K_a^*+K_b^*) & t_1K_b^*(1+K_a)  \\
 t_1(K_a+K_b) & (\epsilon_A-t_2) & t_1K_a(1+K_b^*)  \\
 t_1K_b(1+K_a^*) & t_1K_a^*(1+K_b) & (\epsilon_A-t_2)  
\end{bmatrix}
\nonumber
. 
\end{align}
This demonstrates that the flat band of the kagome lattice is embedded in the bitriangular lattice under the imposed condition for the 
flat band, whose new energy is $\lambda=\epsilon_A-t_2-2t_1$. 
An example for the parameters can be obtained again by performing first-principles calculations with the H $s$ orbital. 
The lattice constant shown in Fig.~\ref{fig:fig2} (d) is set to 7 \AA~to avoid long-range hopping.
We then tune the height of atom D and find that a flat band is revealed when the height reaches 2.995 \AA, as 
shown in Figs.~\ref{fig:fig2} (e) and (f). The tight-binding parameters obtained from the Wannier functions are listed in Table~\ref{table:table1}.

\subsection{Checkerboard lattice}
\label{sec:checkerboard}

\begin{figure}[tbp]
\includegraphics[width=1.00\columnwidth,clip=true,angle=0]{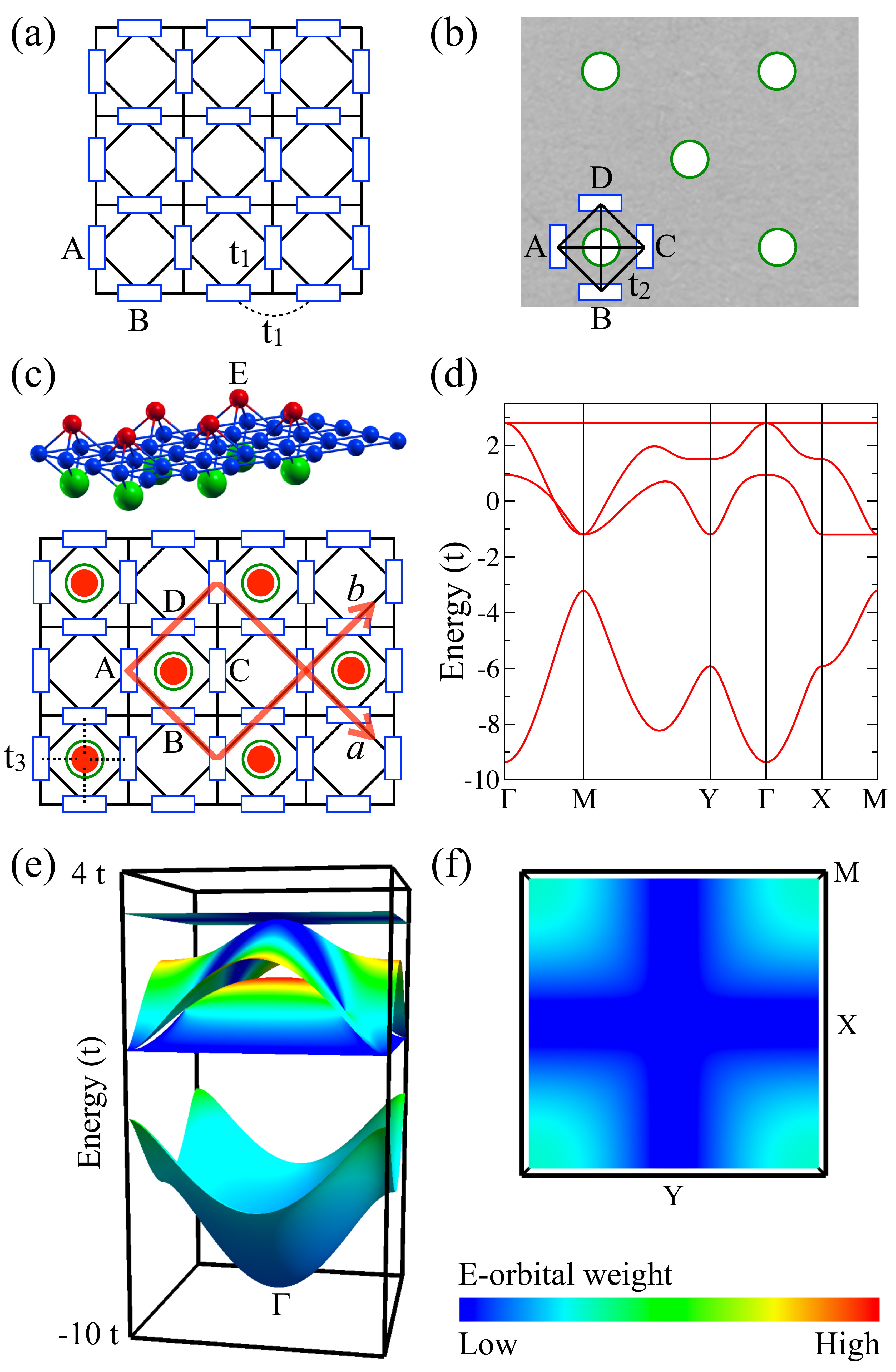}
\caption{(a) Checkerboard lattice with six hopping parameters of the same strength ($t_1$) per site. 
(b) A substrate having protruding atoms (green circles) on the surface. The effect of the presence of protruding atom 
is to enhance an equal amount of hopping strength to the hopping parameters between orbitals A, B, C, and D 
by $t_2$. (c) The checkerboard lattice on the substrate with additional adatoms (red circles).
The orbital at the adatom can hop to orbitals A, B, C, and D via the parameter $t_3$. The supercell is indicated by the red lines. 
(d) Band structure given by the parameters listed in Table~\ref{table:table1} along the same path shown in Fig.~\ref{fig:fig1} (e).
(e) Three-dimensional band structure with E-orbital contribution. (f) Top view of the flat band with E-orbital weight. 
}
\label{fig:fig3}
\end{figure}

The deformation applied to the lattices that host flat bands can also be introduced by the presence of a substrate. 
The third example we will show is the checkerboard lattice. The checkerboard lattice that can give a flat band
is shown in Fig.~\ref{fig:fig3} (a), where an orbital having the $d_{z^2}$ shape is located at each site 
and ordered in a way that both first-neighbor and second-neighbor hopping parameters have the same strength (denoted as $t_1$) 
at a delicate lattice constant. The flat band can be found at the energy of $\epsilon-2t_1$.
By fabricating the checkerboard lattice on the substrate surface having protruding atoms, 
for example, the one shown in Fig.~\ref{fig:fig3} (b),  
some sort of effective hopping has also been introduced to the system.
Following the idea that the system allows for six hopping paths of the same strength per site, we also assume the effect
of the protruding atom is to increase the hopping strength by an amount of $t_2$ for the six surrounding paths as shown in Fig.~\ref{fig:fig3} (b).
The way to eliminate the effect of $t_2$ that has modified the flat band is to add an adatom E in a supercell for 
resulting in a new hopping parameter $t_3$ that can hop to its first-neighbor orbitals A, B, C, and D.  
The new system is indicated by the supercell shown in Fig.~\ref{fig:fig3} (c).

The band structure of the modified system can be obtained by solving the eigenvalue problem:
\begin{widetext}
\begin{align}
\begin{bmatrix}
 \epsilon          & t_2+t_1(1+K_a^*)   & t_2+t_1(K_a^*+K_b^*) & t_2+t_1(1+K_b^*)   & t_3        \\
 t_2+t_1(1+K_a)  & \epsilon             & t_2+t_1(1+K_b^*)     & t_2+t_1(K_a+K_b^*) & t_3        \\
 t_2+t_1(K_a+K_b)& t_2+t_1(1+K_b)     & \epsilon               & t_2+t_1(1+K_a)     & t_3        \\
 t_2+t_1(1+K_b)  & t_2+t_1(K_a^*+K_b) & t_2+t_1(1+K_a^*)     & \epsilon             & t_3        \\
 t_3               & t_3                  & t_3                    & t_3                  & \epsilon_E 
\end{bmatrix}
\begin{bmatrix}
C_A \\
C_B \\
C_C \\
C_D \\
C_E 
\end{bmatrix}
=  \lambda
\begin{bmatrix}
C_A \\
C_B \\
C_C \\
C_D \\
C_E
\end{bmatrix}
\label{eq:eq12}
.
\end{align}
\end{widetext}
To reveal the embedded checkerboard ingredient that gives the flat band in the modified system, the last row of Eq.~\ref{eq:eq12}
can be first rewritten as
{\small
\begin{align}
t_2 C_A + t_2 C_B + t_2 C_C + t_2 C_D + t_2(\epsilon_E-\lambda)/t_3 C_E = 0. 
\end{align}
}
For the cancellation of $t_2$ and $t_3$ in the first four rows of Eq.~\ref{eq:eq12}, we have reached the same needed condition:
\begin{align}
t_2(\epsilon_E-\lambda)/t_3 = t_3, \label{eq:eq14}
\end{align}
which can be realized by, for example, tuning the height of atom E. The new energy eigenvalue is $\lambda=\epsilon-t_2-2t_1$.
One example of the tight-binding parameters is given in Table~\ref{table:table1}, and the band structure is shown in Figs.~\ref{fig:fig3} (d) and (e).
We emphasize again that this embedding is only possible
because of the flat-band nature in the original checkerboard lattice that gives a $k$-independent constant 
for each parameter in Eq.~\ref{eq:eq14}. We further note that under the condition of Eq.~\ref{eq:eq14}, 
Eq.~\ref{eq:eq12} can be reduced back to the equations expressed by the orbitals in
the original unit cell for realizing the flat band even though the supercell is needed in the presence of the adatom and substrate.

\subsection{Discussion}

The eigenvector coefficients of the flat band in each of the illustrated deformed systems are the same as those in the original system
at every individual $k$ point in the entire Brillouin zone regardless of the renormalization factor due to the presence of the adatom. 
In Figs.~\ref{fig:fig1} (g), \ref{fig:fig2} (g), and \ref{fig:fig3} (f), the contributions of the adatoms to keep the bands highly degenerate in
energy are plotted for the Lieb, kagome, and checkerboard lattices, respectively. Note that the major adatom weight is not 
on the flat bands; otherwise, the flat bands just belong to trivial isolated states constructed by the adatom orbitals with $t_3\sim 0$.
As already discussed for the three examples, Eq.~\ref{eq:eq9} and Eq.~\ref{eq:eq14} cannot be satisfied by just tuning the property of an adatom without the help of the hidden symmetry
that can give rise to a flat band because of the $k$-dependent eigenvalue $\lambda(\vec{k})$. Consequently,
many interesting physical phenomena associated with the flat bands, such as ferromagnetism, superconductivity, and fractional quantum Hall effects
\cite{PhysRevLett.106.236802,PhysRevLett.106.236803,PhysRevLett.106.236804,PhysRevLett.107.146803,PhysRevLett.99.070401,PhysRevB.77.235107,
PhysRevB.82.184502,PhysRevLett.84.143,peotta2015superfluidity,PhysRevB.81.113104,PhysRevB.85.205128,PhysRevLett.114.245503,
PhysRevLett.114.245504,PhysRevB.93.075126,Xia2016,Zong2016,PhysRevA.80.063603,PhysRevB.84.115136,PhysRevA.83.063601,PhysRevB.99.045107},
are expected to be found in more materials beyond currently investigated structures\cite{C8TA02555J,C8NR08479C,su2018prediction,jiang2019lieb,PhysRevLett.121.096401,Marchenkoeaau0059,Lieaau4511}.

Generally speaking, the new systems in which the recognized flat bands are embedded cannot be considered as the original systems 
since the new band structures would deviate from the original ones due to the non-zero $t_2$ and $t_3$. 
However, the embedded ingredients can still be observed in the wave functions that give the flat bands in the new systems
at certain $k$ points, where the adatoms become invisible as shown by the zero (blue) weight in Figs.~\ref{fig:fig1} (g), \ref{fig:fig2} (g), and \ref{fig:fig3} (f)
for the Lieb, kagome, and checkerboard lattices, respectively.  
For example, one can confirm that the eigenvector coefficients at $\Gamma$ point 
between the original and the deformed systems are exactly the same in the introduced Lieb, kagome, or checkerboard lattice. 
Even though additional orbitals are added into those systems through the adatoms, their eigenvector coefficients at $\Gamma$ point are zero 
once the discussed condition is satisfied. In such a case, the flat-band eigenstates in the two systems are identical except the real-space distortion 
due to the introduced displacement. Finally, it is worth mentioning that the exotic feature in the kagome lattice, namely the coexistence of one flat band 
and two Dirac bands, can still be found with the presence of the adatom, as shown in Fig.~\ref{fig:fig2} (f), where the blue parts of the bands provide 
a good example for the equivalency between the kagome lattice and the bitriangular lattice\cite{PhysRevB.99.100404}. 

\section{Conclusion}

We have shown that the flat bands given by the well-known lattices, such as the Lieb, kagome, and checkerboard lattices, 
can be embedded in the new structures that cannot be recognized as the original ones, indicating that interesting flat-band physics can be
realized in a larger amount of materials. Although the new Hamiltonian cannot be obtained by
a unitary transformation from the original system, such embedding is mathematically and physically exact in the sense that they satisfy the same
eigenvalue equations for revealing the flat bands, and the introduced additional tight-binding parameters to the original system  
cannot affect the flat-band energy eigenstates at the $\Gamma$ point.  
This hidden mechanism for realizing the flat bands cannot be directly predicted by the line graphs because the embedded ingredients are hidden 
in different structures. Such finding opens a new avenue for designing nearly flat bands around the imposed condition by
selecting different species and/or heights of the adatoms in a variety of electronic, photonic, and other interesting  materials.

\begin{acknowledgments}
We are grateful for the use of supercomputers at JAIST.
This work was supported by Priority Issue (creation of new functional devices and high-performance materials to 
support next-generation industries) to be tackled by using Post `K' Computer, Ministry of Education, Culture, Sports, Science and Technology (MEXT), Japan.
A part of this work was carried out using the facilities in JAIST, supported by Nanotechnology Platform Program (Molecule and Material Synthesis) of
the Ministry of Education, Culture, Sports, Science and Technology (MEXT), Japan.
\end{acknowledgments}

\bibliography{refs}

\end{document}